\documentclass[letterpaper, 10 pt, conference]{ieeeconf}  

\IEEEoverridecommandlockouts                              

\usepackage{graphics}
\usepackage{graphicx}

\title{\LARGE \bf
Feel the Presence: The Effects of Haptic Sensation on VR-Based Human-Robot Interaction}

\author{Xinyan Yu$^{1}$, Marius Hoggenmüller$^{1}$, Tram Thi Minh Tran$^{1}$ and Martin Tomitsch$^{2}$
\thanks{$^{1}$Design Lab, Sydney School of Architecture, Design and Planning, The University of Sydney, Australia
        {\tt\small contact author: xinyan.yu@sydney.edu.au}}%
\thanks{$^{2}$ Transdisciplinary School, University of Technology Sydney, Australia 
        {\tt\small}}%
}

\begin{document}

\maketitle
\thispagestyle{empty}
\pagestyle{empty}

\begin{abstract}

Virtual reality (VR) has been increasingly utilised as a simulation tool for human-robot interaction (HRI) studies due to its ability to facilitate fast and flexible prototyping. Despite efforts to achieve high validity in VR studies, haptic sensation, an essential sensory modality for perception and a critical factor in enhancing VR realism, is often absent from these experiments. 
Studying an interactive robot help-seeking scenario, we used a VR simulation with haptic gloves that provide highly realistic tactile and force feedback to examine the effects of haptic sensation on VR-based HRI.
We compared participants' sense of presence and their assessments of the robot to a traditional setup using hand controllers. Our results indicate that haptic sensation enhanced participants' social and self-presence in VR and fostered more diverse and natural bodily engagement. Additionally, haptic sensations significantly influenced participants’ affective-related perceptions of the robot. Our study provides insights to guide HRI researchers in building VR-based simulations that better align with their study contexts and objectives.

\end{abstract}








\section{Introduction}

The integration of robots into everyday life requires the development of effective interaction strategies, which in turn necessitates appropriate methods for prototyping and evaluation~\cite{Woods2006MethodologicalIssues, Riek2012WoZHRI,Zamfirescu2021Fake}. However, prototyping and evaluation interactions with real robots present challenges, including high development costs, potential risks to participants, and difficulties in replicating real-world contexts. To address these challenges, researchers increasingly employ virtual reality (VR) as a prototyping tool to simulate human-robot interactions (HRI) in controlled lab settings~\cite{Miner1994VRrobots, Khastgir2015VRDrive, Marius2021PrototypingAV}.

Despite its potential, ensuring the validity of a VR study is crucial to achieving reliable results that can be translated to real-world settings~\cite{Wijnen2020ReplicatingHRI}. 
While current simulation methods for HRI experiments strive to recreate realistic visual and auditory experiences~\cite{Marius2021PrototypingAV}, haptic sensation---one of the most important sensory modalities for perception~\cite{SRINIVASAN1997Haptics} that enhances VR realism~\cite{Hoffman1998TouchRealisim}---is often absent from the experience. In most HRI studies, even when contact-based physical interaction strategies are tested, the study setups typically rely on hand controllers or hand tracking to facilitate these interactions~\cite{Ortenzi2022PassMeTheTool,Higgins2024CollaborativeBuildingTask}, without enabling participants to experience tactile and force feedback when simulating embodied interactions with robots. 

Haptic sensation has been shown to play a crucial role in shaping how people behave~\cite{Ebrahimi2016Haptic,  Jacucci2024HapticsReview}, and interact with virtual human avatars in VR during social interactions~\cite{krogmeier2023human,Jeremy2003Interpersonal, Venkatesan2023CrowdVR, Giannopoulos2008HapticSocialVR} (
e.g., influencing proxemic behaviour~\cite{Jeremy2003Interpersonal}, intensifying emotional engagement~\cite{Venkatesan2023CrowdVR,Ahmed2016TouchMe}). However, interactions with robots differ fundamentally from those with humans due to differences in agency, embodiment, and social expectations~\cite{FONG2003Survey}. Thus, how haptic sensation impacts people's perception of robots and their interactions with them in VR remains unknown. Beyond social perception and interaction, the lack of realistic haptic simulation can also greatly limit the effectiveness of VR for evaluating a product's usability and user experience~\cite{Bruno2010MixedPrototyping}. This further emphasises the importance of studying the impact of replicating haptic sensations on people's assessments of robots in VR-based studies.

Advances in haptic technology have enabled the direct touch and manipulation of computer-generated objects in VR in ways that evoke a compelling sense of tactile realness~\cite{SREELAKSHMI2017HapticTechnology,Caeiro2021GlovesReview, Sada2918HapticSerpent}. Industrial-grade tools such as HaptX Gloves\footnote{https://haptx.com/} can provide high-quality haptic feedback through microfluidic technology, simulating force feedback (e.g., hardness, weight, resistance) and tactile feedback (e.g., texture, smoothness, friction) to enhance realism in virtual interactions. HaptX Gloves have been integrated into Volkswagen's vehicle design and training VR platform to support in-vehicle prototyping and evaluation~\cite{Stamer2020HapticBenifit}. In addition, a growing range of affordable and diverse commercial smart gloves has made haptic technology increasingly accessible to researchers~\cite{Caeiro2021GlovesReview}.

Building on these advancements, we explore the potential of highly realistic haptic feedback to enhance VR-based HRI experiments by incorporating HaptX Gloves into a VR setup, enabling participants to physically interact with virtual representations of robots and their surrounding environment. To examine the effects of haptic simulation using haptic gloves on participants' sense of presence and their assessments of robots, we replicated a robot help-seeking study that we previously conducted~\cite{Yu2024Playful}. In the replication study presented here, we compared a haptic gloves setup, which provides participants with tactile and force feedback and allows them to directly interact with the virtual robot using their hands, with the original setup that used hand controllers for interaction.
Our work makes the following contributions: (1) We present a VR-based HRI study setup that incorporates highly realistic haptic gloves, enabling haptic sensations during physical interactions with virtual robots; (2) We provide empirical evidence on how such a setup influences participants' sense of presence, bodily engagement, and affective perceptions of the robot, offering practical guidance for researchers building VR-based HRI studies.

\section{Related Work}

\subsection{VR as a prototyping tool for HRI}
VR as a prototyping tool for HRI has demonstrated its ability to effectively simulate real-world human-robot interactions. For instance, Villani et al.~\cite{Villani2018VRforComplexHRI} found comparable results in task performance and evaluations when comparing interactions with real robots to those in VR within a human-robot collaborative task setting. Similarly, Sadka et al.~\cite{Sadka2020} observed that social interpretations of a non-humanoid robot's gestures in VR closely resemble those in the real world. 

However, notable differences remain in how participants interact with and assess robots in VR compared to real-world settings. For instance, Wijnen et al.~\cite{Wijnen2020ReplicatingHRI} replicated a study by Kahn et al.~\cite{Kahn2015secret} on people's secret-keeping behaviour with robotic and human tour guides in VR. The replication VR study found that participants were more likely to keep secrets for the non-social robot, contradicting the original study's results. In a human-robot collaborative building task study, Higgins et al.~\cite{Higgins2024CollaborativeBuildingTask} reported that each attempt by participants to complete the task took significantly longer in VR. While this study did not directly examine the impact of haptic sensation, the authors attributed the performance differences to difficulties in depth perception in VR, a perception in which haptic sensation plays a key role~\cite{Makin2019Tactile}. Despite the recognised importance of haptic sensation for enhancing VR realism, VR-based HRI studies, even those involving contact-based interactions, rely on hand controllers or hand tracking~\cite{Ortenzi2022PassMeTheTool,Higgins2024CollaborativeBuildingTask} as participant interaction method, limiting participants from experiencing tactile and force feedback with virtual robots. However, investigations into how a VR setup that enables realistic haptic sensations affects participants' behaviours and assessments of robots in VR-based HRI studies remain limited.
\subsection{Importance of haptics in VR}
Haptic sensation is essential for enhancing immersion and realism in VR environments~\cite{Bystrom1999SoP,Wijnand2000Presence}. To enable touch-based interactions, various devices have been developed to simulate haptic sensations, allowing users to perceive tactile and force feedback when engaging with virtual objects, with most focusing on the hands as the primary modality for interaction~\cite{Perret2018HapticGloves,Heather2018Haptics,Pacchierotti2017Wearable}. Integrating haptic feedback into VR simulations has demonstrated benefits across multiple domains, including improving performance in VR surgical training~\cite{GANI20221SurgicalTrainning}, enhancing precision in fine manipulation tasks such as drawing~\cite{Richard2021Drawing}, reducing execution time for tasks like throwing and stacking~\cite{Kreimeier2019HapticTaskBase}, or improving the remote manipulation of robots~\cite{Brogi2024Avatar, Ni2017Haptic}. Beyond task performance, haptic feedback has been shown to shape social interactions in VR, influencing how people engage with virtual human avatars~\cite{krogmeier2023human,Jeremy2003Interpersonal,bailenson2008virtual,Giannopoulos2008HapticSocialVR, Venkatesan2023CrowdVR}. For example, Jeremy et al.~\cite{Jeremy2003Interpersonal} found that participants maintain larger interpersonal distances from virtual avatars in VR compared to real-world interactions, due to the absence of tactile haptic contact with virtual representations. Haptic sensation has also been shown to enhance the sense of social presence when interacting with another human avatar in shared virtual environments~\cite{Giannopoulos2008HapticSocialVR}. While these studies focus on task performance or interactions with virtual avatars of another human, the impact of haptic sensation on people's perception and behaviour toward virtual robots, given their distinct agency, embodiment, and social expectations~\cite{FONG2003Survey}, remains unknown. 

\subsection{Haptic sensation in VR-based prototyping and evaluation}

VR has been widely adopted as a prototyping tool for studying interactions with interactive systems, such as robots~\cite{Miner1994VRrobots}, autonomous vehicles~\cite{Khastgir2015VRDrive, Marius2021PrototypingAV} and smart devices~\cite{Voit2019ResearchMethod}, reducing complexity, costs, and risks in development and evaluation. Previous work has shown that haptic sensation plays a critical role in facilitating effective assessments in VR-based prototyping and evaluation. Schölkopf et al.~\cite{SCHOLKOPF2021Hapticfeedbackismoreimportant} highlighted that the absence of haptic feedback severely diminishes user experience ratings, emphasising its importance for meaningful user experience assessments in VR. Similarly, Stamer et al.~\cite{Stamer2020HapticBenifit} investigated high-fidelity haptic feedback in user in-vehicle experience evaluation and found that participants recognised interactions significantly faster and more accurately when haptic feedback was present compared to when it was absent. While these studies demonstrate the value of realistic haptic simulation in VR-based prototyping with traditional user interfaces, its role in supporting the assessment of physically and socially embodied robots remains largely unexplored.

\section{Methodology}
\subsection{Study design}

\begin{figure*}[t]
    \centering
    \includegraphics[width=1\textwidth]{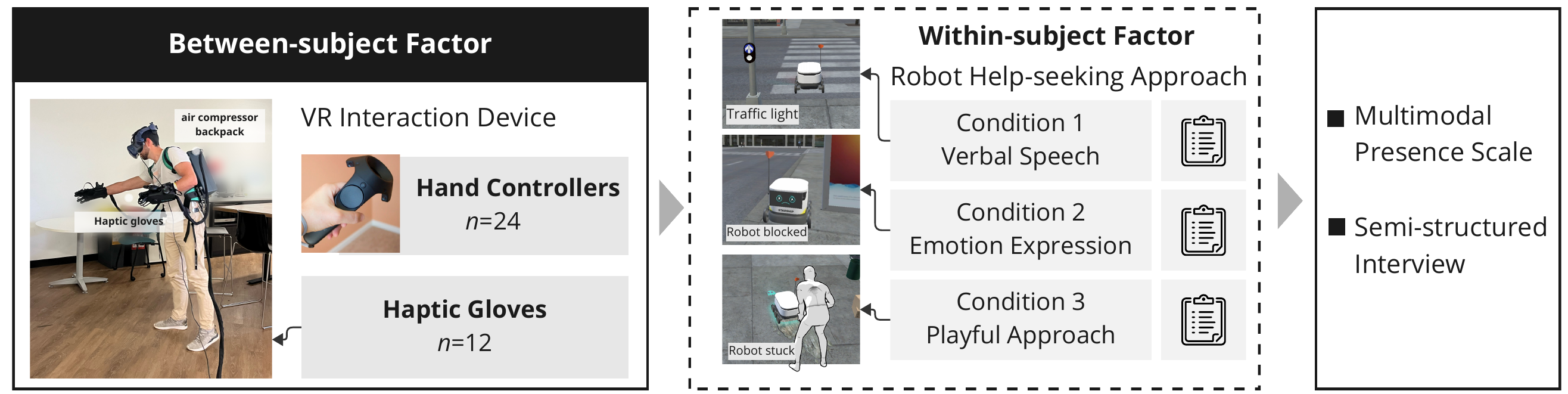}
    \caption{Study Design}
    \label{StudyDesign}
\end{figure*}

This study replicates our previous study of how urban robots that get stuck seek help from bystanders~\cite{Yu2024Playful}. While maintaining the same study procedure and data collection methods as the original, we replaced the hand controllers in the original setup with haptic gloves to investigate the effects of haptic simulation. The interaction device (i.e., hand controllers or haptic gloves), constitutes the between-subject factor and is the primary focus of our investigation in this paper. To ensure methodological consistency with the original study, participants within each group experienced multiple robot help-seeking methods (i.e., speech, emotional expression, and playful help-seeking, see Fig.~\ref{StudyDesign}), forming the within-subject factor. 
For speech help-seeking, the robot emitted a voice prompt, such as \emph{`Please help me press the traffic light button, I am going to be late.'} For emotional help-seeking, the robot displayed a sad facial expression. For playful help-seeking, the robot used an interactive ground projection inspired by games to elicit helping behaviours\footnote{For more details on these help-seeking approaches, please refer to the original study~\cite{Yu2024Playful}.}. While these within-subject conditions are not the primary focus of this study, their inclusion maintains alignment with the design and scope of the original investigation.

Twenty-four participants from the original study formed the \emph{Hand Controllers} group (A1-A24, aged 18 to 74 years, with most (n=17) between 25 and 34 years), while 12 participants completed our study using haptic gloves, forming the \emph{Haptic Gloves} group (B1-B12, aged 18 to 34 years, with most (n=8) also between 25 and 34 years). Since the primary focus of this study was on between-subject group effects rather than comparing different help-seeking methods, the help-seeking methods were treated as repeated measures, allowing us to achieve adequate statistical power with a smaller sample size.


\subsection{Study procedure}
Upon arrival, participants received a brief introduction to the study's background and procedure, followed by instructions on the VR devices and their use. Before the experiment, they went through a familiarisation session, practising walking and interacting with virtual objects in VR (e.g., pushing a rack on wheels) using either hand controllers or haptic gloves. During the experiment, participants followed the same procedure as the original study, experiencing three within-subject conditions: the robot's help-seeking methods of speech, emotional expression, and playful help-seeking, in a counterbalanced order. The study was situated in a virtual urban environment, where robots seek assistance from random pedestrians. Participants experienced each robot help-seeking methods multiple times across three scenarios (see Fig.~\ref{StudyDesign}): \emph{Robot blocked}, where the robot is obstructed requires participants to make way; \emph{Traffic light}, where the robot requires participants to press the traffic light button for it; and \emph{Robot stuck}, where the robot is immobilised and depends on participants to help it out of its stuck position. 
Participants were not informed about the robot’s intention to seek help and were instructed to respond spontaneously to its cues.
After each help-seeking condition, participants removed the headset to fill out several standardised questionnaires that assessed their perceptions of the robot. Upon completing all trials, they filled out an additional questionnaire on their sense of presence in VR. We also conducted a post-study semi-structured interview to gain in-depth insights into their experiences. Each session took approximately 60 minutes.

\subsection{Study apparatus and implementation}

The VR simulation was developed in Unity3D\footnote{https://unity.com/} and deployed on an HTC Vive headset\footnote{https://www.vive.com/}. The experiment was conducted in a 5x5-metre open floor space, allowing participants to physically walk within the simulated urban space. Participants in the original study used the default HTC Vive controllers, while those in our \emph{Haptic Gloves} group interacted with the virtual environment directly using HaptX Gloves (see Fig.~\ref{StudyDesign}). 
The HaptX Gloves used in our study provide high-fidelity tactile feedback on the fingers and palms through microfluidic actuators, which inflate and deflate to create localised pressure, simulating the texture details of virtual objects. They also provide force feedback via a mechanical resistance system that restricts finger movement to render varying levels of stiffness, allowing participants to perceive the weight and rigidity of objects in VR. The HaptX system operates via a compressed air-powered setup, consisting of an air compressor backpack weighing approximately 9 kg and a pair of gloves (see Fig.~\ref{StudyDesign}, left). The backpack supplies compressed air through pneumatic tubing, driving the actuators in the gloves to deliver real-time tactile feedback. Participants wore the backpack and gloves, allowing them to directly manipulate virtual objects with their hands and feel them as if they were physical. This setup enabled realistic interactions with all VR objects within participants’ reach, including the robot (e.g., pushing the robot) and environmental elements (e.g., pressing a traffic light button).

\subsection{Data collection}

\textit{Sense of presence}: Participants' sense of presence in VR was evaluated using the same Multimodal Presence Scale (MPS)~\cite{MAKRANSKY2017MPS} from the original study, encompassing three dimensions: physical, social, and self-presence.

\textit{Perceptions of the robot}: To examine how haptic sensations influence participants' assessments of the robot in VR, we evaluated their perceptions across several dimensions aligned with the original study. The Robotic Social Attributes Scale (RoSAS)~\cite{Carpinella2017RoSAS} assessed participants’ perceptions of the robot, including warmth, competence, and discomfort. The System Acceptance Scale~\cite{VANDERLAAN19971} measured participants’ acceptance towards the robot, while the likeability subscale from the Godspeed Questionnaire~\cite{bartneck2009GodSpeed} evaluated perceived likeability. Trust was measured using the trust scale~\cite{McAllister1995AffectAC}, which consists of two subscales: cognitive trust and affective trust. Lastly, the short version of the User Experience Questionnaire (UEQ-S)~\cite{Martin2017UEQ} captured participants’ overall user experience. All measures used a 7-point Likert scale.

\textit{Qualitative data}: We took observation notes of participants’ behaviours during the study. In addition, we conducted a semi-structured interview with participants after they completed all experimental tasks, during which we asked questions about their overall experience in VR and their perceptions of the robots.

\subsection{Data analysis}
\emph{Quantitative analysis}: 
We first assessed the internal reliability of all multi-item scales by calculating Cronbach’s alpha. For the MPS, physical presence showed acceptable reliability ($\alpha=0.747$), and self presence demonstrated excellent reliability ($\alpha=0.892$). The social presence subscale had questionable reliability ($\alpha = 0.674$). Following advice on Cronbach’s alpha in ~\cite{tavakol2011making}, we removed items with correlations below 0.25 for following analysis, which improved reliability to an acceptable level ($\alpha = 0.710$). The likeability scale received excellent reliability ($\alpha$=0.923). The trust scales showed excellent reliability for affective trust ($\alpha$=0.930) and good reliability for cognitive trust ($\alpha$=0.823). The System Acceptance Scale showed good reliability ($\alpha$=0.887). For the RoSAS, competence and warmth showed 
good ($\alpha$=0.867) and excellent ($\alpha$=0.900) reliability, respectively, while discomfort had acceptable reliability ($\alpha$=0.761). Lastly, the UEQ-S scales showed good reliability for pragmatic quality ($\alpha$=0.806) and excellent reliability for hedonic quality ($\alpha$=0.904).

To compare sense of presence between groups (hand controllers vs. haptic gloves), we conducted independent samples t-tests for each MPS subscale. Given the unequal sample sizes between groups, we assessed variance homogeneity using Levene’s Test. When equal variances were violated (p < 0.05), Welch’s t-test was applied to account for heterogeneity. Additionally, effect sizes (Cohen’s $d$) were calculated to provide insight into the practical significance of group differences.

For participants' robot assessments, we applied a Linear Mixed Model (LMM) to account for repeated measures across the within-subject factor (Conditions, i.e., robot help-seeking approaches). The between-subject factor (Groups, i.e., hand controllers vs. haptic gloves) was specified as the primary fixed effect of interest. The interaction effect between Group and Condition was included to examine whether the impact of Conditions varied across Groups. Participants were treated as a random effect to account for inter-individual differences. 
The LMM calculated estimated marginal means for fixed effect and also accounted for unequal sample sizes and modelled dependencies between repeated measures using a Compound Symmetry covariance structure. Effect sizes (Cohen’s $d$) were calculated to assess practical significance. 

{\emph{Interviews:}} We employed an inductive semantic analysis~\cite{thematicAnalysis} to identify patterns in participants' comments that could supplement our understanding of the quantitative results. Interviews were transcribed and analysed by the first author.

\section{Results}
This section begins with observations of participants' behaviours and related interview comments. We then present differences in sense of presence between groups, supported by interview insights to contextualise the results. Finally, we report statistical findings on participants' assessments of robots, including main group effects and interaction effects.

\subsection{General observations}
The robot received help at similar rates across both groups. In \emph{Haptic Gloves} group (3 × 3 × 12 trials), help was provided in 104 trials (96.3\%). In the \emph{Hand Controllers} group (3 × 3 × 24 trials), help was provided in 209 trials (96.8\%). Despite similar helping decisions, participants in \emph{Haptic Gloves} group demonstrated more natural and diverse full-body engagement during the study, including tapping the robot with hands (n=4), using their feet to interact with it (n=3), dragging the robot's flagpole (n=2), or giving it a high-five (n=1). In contrast, these spontaneous and varied bodily engagements were rarely observed in \emph{Hand Controllers} group, where only one instance of tapping the robot occurred.

Post-study interviews provide insights into these behavioural differences. Five participants attributed their more frequent and diverse bodily engagement to the absence of controller constraints, referencing their previous experiences using hand controllers in other VR settings. They noted that haptic gloves allowed free-hand interaction without the need to hold a controller, enabling more direct and natural actions that closely aligned with real-life interactions. B5 and B9 stated that they would not have performed such actions with hand controllers and explicitly cited this constraint, explaining that holding controllers restricted their ability to engage in these interactions: \emph{‘I don’t think I would (high-five the robot) because I am not able to. With the controller, you have to hold it’} (B9). In contrast, B2 attributed their behaviour specifically to the increased tangibility of the robot in VR, made possible by the tactile and force feedback provided by the haptic gloves. This enhanced the robot's physical presence and led to the participant's greater emotional investment, ultimately prompting B2 to kick the robot when it failed to respond.

In addition, B2 and B4 expressed a heightened sense of full-body engagement due to the high level of immersion, with B2 describing it as \emph{‘my whole body got activated’}. In contrast, A8 and A11 from \emph{Hand Controllers} group expressed an intention to use their feet to interact with the robot (e.g., to help it get unstuck) but did not act on it, citing uncertainty about whether such actions were \emph{‘allowed.’} Moreover, six participants from the \emph{Hand Controllers} group reported difficulties with distance perception in VR, a challenge mentioned only once in the \emph{Haptic Gloves} group. This limitation likely hindered participants in the \emph{Hand Controllers} group from engaging in more embodied and exploratory behaviours.

\subsection{Sense of presence}

\begin{figure*}[t]
\begin{center}
\includegraphics[width=0.9\textwidth]{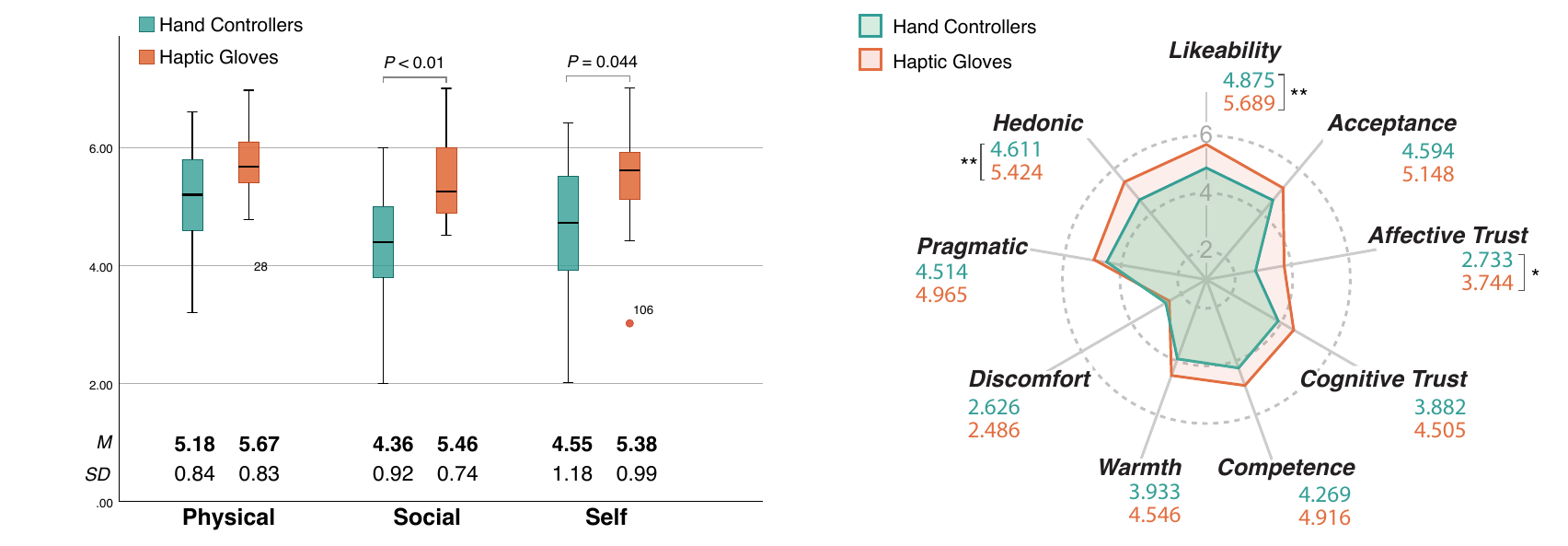}
\end{center}
\caption{Quantitative Results. Left: Box plot of Multimodal Presence Scale (M: Mean, SD: Standard Deviations). Right: Participants’ assessments of the robot based on estimated marginal means (*: $p < 0.05$, **: $p < 0.01$, ***: $p < 0.001$).}\label{Results}
\end{figure*}

\textit{MPS}: Descriptive analysis results indicated that participants in the \emph{Haptic Gloves} group reported a higher sense of presence across all subscales compared to the \emph{Hand Controllers} group (see Fig.~\ref{Results}). Independent samples t-tests showed no significant difference in physical presence ($t(34) = -1.662, p = 0.106$), but significantly higher scores for the \emph{Haptic Gloves} in social presence ($t(34) = -3.601, p < 0.001$), and self presence ( $t(34) = -2.094, p = 0.044$). The effect sizes (Cohen’s $d = 0.86$ for social presence and $d = 1.13$ for self presence) indicate large effects, suggesting that the observed differences are practically meaningful despite the unequal sample sizes between groups.

\textit{Qualitative feedback}: While all participants in \emph{Haptic Gloves} group agreed that haptic feedback enhanced their sense of presence, eight noted that discrepancies with real-world haptic experience inevitably exist, such as the gloves' inability to simulate friction when touching objects and the insufficient force resistance when pushing the robot, failing to accurately reflect its real-world weight. Additionally, five participants noted that the device's substantial weight and its tethered nature of cable connection imposed limitations on their freedom of action. These limitations may have contributed to the absence of a significant difference in physical presence between groups.

Despite these limitations, four participants indicated that the natural interaction enabled by the haptic gloves led to a higher level of engagement in the virtual scene compared to their previous experiences in VR without haptic gloves. Some attributed this engagement to the greater actual physical effort required in the haptic glove setup (n=2). B2 additionally remarked that the experience felt more emotionally engaging due to the sensations being closer to real life. In contrast, A2 from \emph{Hand Controllers} group likened pushing the robot by holding controllers to \emph{‘using a magical wand’}, suggesting lower engagement and seriousness compared to the haptic gloves experience.

The interviews also reflected how being able to physically feel the robot influenced participants’ perceptions of the robot's presence. B4 noted that physically touching the robot made it \emph{‘more convincing that the robot truly existed,’} describing it as \emph{‘something real in front of me.’} Conversely, A2 and A5 from \emph{Hand Controllers} group remarked that the lack of haptic feedback made the interaction feel like \emph{‘pushing the air,’}~(A5) leading them to realise \emph{‘the robot is not real.’}

Four participants in \emph{Haptic Gloves} group noted that the gloves gave them a sense of greater control over their body, with B12 stating, \emph{‘I have some power about using my hand or my body,’} and B11 describing \emph{‘my hand and my virtual hand become one.’} In contrast, three participants from \emph{Hand Controllers} group reported feeling disconnected between their own body and the virtual one, stating \emph{‘the hands didn’t really feel like my hands.’}~(A18). These differences in experiencing embodiment could potentially be related to the higher self-presence scores observed in the \emph{Haptic Gloves} group.

\subsection{Participants assessments of the robot}

\textit{Likeability}:
The \emph{Haptic Gloves} group reported higher likeability scores than \emph{Hand Controllers}, with a significant mean difference ($M_{diff} = 0.814, SE = 0.259, p = .004$, 95\% CI [0.287, 1.341]) (Fig.~\ref{Results}, right). The effect size (Cohen’s $d = 0.75$) suggests a moderate-to-large practical effect. The Group $\times$ Condition interaction effect was non-significant, $F(2, 68) = 0.537, p = .587$, indicating differences in likeability scores across conditions were consistent between groups.

\textit{Acceptance}:
The \emph{Haptic Gloves} group reported higher mean acceptance towards the robot compared to the \emph{Hand Controllers} group, but this difference was not significant ($M_{diff} = 0.554, SE = 0.291, p = .067$, 95\% CI [$-0.036$, $1.144$]). The Group $\times$ Condition interaction effect was also non-significant ($F(2, 68) = 0.416, p = .661$).

\textit{Trust}: 
For affective trust, \emph{Haptic Gloves} group reported higher scores than \emph{Hand Controllers} group, with a significant mean difference ($M_{diff} = 1.011, SE = 0.485, p = .037$, 95\% CI [$0.064$, $1.958$]). The effect size (Cohen’s $d = 0.67$) was moderate. For cognitive trust, the \emph{Haptic Gloves} group also reported higher scores, but this difference was not significant ($M_{diff} = 0.623, SE = 0.375, p = .107$, 95\% CI [$-0.139$, $1.385$]). The Group $\times$ Condition interaction effect was non-significant for both affective ($F(2, 68) = 0.627, p = .537$) and cognitive trust ($F(2, 68) = 1.850, p = .165$), indicating that differences in trust scores across conditions were consistent between groups.

\textit{Robotic social attributes}: 
Although \emph{Haptic Gloves} group reported slightly higher competence ratings than \emph{Hand Controllers} group, the difference was not significant, $M_{diff} = 0.647, SE = 0.309, p = .069$, 95\% CI [$-0.024$, $1.317$]. Similarly, warmth ratings were slightly higher in \emph{Haptic Gloves} group, but this difference was not significant($M_{diff} = 0.613, SE = 0.378, p = .119$, 95\% CI [$-0.154$, $1.380$]). Discomfort ratings were comparable between the two groups, with no significant main effect of Group, ($M_{diff} = 0.140, SE = 0.257, p = .590$, 95\% CI [$-0.380$, $0.660$]). No significant Group $\times$ Condition interaction effects were found across all three subscales: Competence, $F(2, 68) = 0.592, p = .556$; Warmth, $F(2, 68) = 0.236, p = .790$; and Discomfort, $F(2, 68) = 0.238, p = .789$.

\textit{User experience}: Pragmatic experience quality ratings were slightly higher in \emph{Haptic Gloves} group compared to \emph{Hand Controllers}, but this difference was not significant ($M_{diff} = 0.451, SE = 0.341, p = .195$, 95\% CI [$-0.242$, $1.144$]). However, hedonic experience quality was significantly higher in the \emph{Haptic Gloves} group ($M_{diff} = 0.813, SE = 0.267, p = .004$, 95\% CI [$0.275$, $1.351$]), with a moderate-to-large effect size (Cohen’s $d = 0.75$). No Group $\times$ Condition interaction effects were found for either subscale—pragmatic quality, $F(2, 68) = 0.227, p = .798$, or hedonic quality, $F(2, 68) = 0.120, p = .887$. 

\section{Discussion}

\subsection{Haptic simulation facilitates spontaneous bodily engagement}
Our study results indicate that enabling haptic interaction in VR-based HRI study using haptic gloves significantly enhances participants' self-presence, defined as the feeling of connection to their virtual body, emotions, and identity~\cite{Lee2006Tactile}. The more frequent and diverse natural bodily engagement observed among participants using haptic devices further highlights that heightened self-presence fosters more natural and a wider range of bodily interactions beyond task-focused behaviours.Spontaneous bodily engagement becomes increasingly critical in VR-based HRI studies as robots transition from static, controlled laboratory settings to everyday environments~\cite{Brown2024Trash,Yu2024Wild}, where interactions involve not only human collaborators who engage in predefined tasks but also casual bystanders. This shift renders human-robot interactions less predictable and more situated and emergent~\cite{Putten2020Forgotten,Xinyan2024CasualCollaboration}, underscoring the need for VR-based HRI experiments to move beyond curated tasks towards capturing spontaneous interactions in dynamic, contextually diverse environments. In this context, simulating haptic sensations in VR plays a pivotal role in rendering a more realistic sense of embodiment that is necessary to effectively study these emergent human-robot interactions. 

\subsection{The relevance of haptic simulation for affective-related assessments}

We did not find a significant interaction effect between Groups and Conditions in any assessment, indicating that the observed differences between robot help-seeking methods are consistent and not influenced by the type of VR interaction devices used. In contrast, Voit et al.~\cite{Voit2019ResearchMethod}'s study on comparing research methods to evaluate smart artefacts identified a significant interaction effect between the evaluation methods and the investigated artefacts, suggesting potential contradictions in artefacts comparison results when different evaluation methods are employed.
The absence of an interaction effect in our study could be attributed to the fact that the inclusion of haptic sensations did not fundamentally alter the evaluation method, as the study remained within a VR setup. This validates VR, even without the inclusion of haptic sensations, as a robust tool for comparative studies in HRI experiments, ensuring that concept comparison results remain consistent regardless of whether haptic sensations are included. 

However, main effects of Groups did exist for scales more closely related to affective perception, including likeability and affective trust. The ability to engage in direct physical interaction with the robot resulted in higher assessments of these affective dimensions (see Fig.~\ref{Results}, right). This could be attributed to two factors. First, our results indicate that participants using haptic gloves experienced enhanced social presence in VR, a sense of being together with another in virtual environments~\cite{Biocca2003SocialPresece}. This aligns with prior work showing that haptic feedback can enhance social presence during interactions with human avatars in VR~\cite{Giannopoulos2008HapticSocialVR}. Higher social presence fosters authentic and nuanced social interactions~\cite{Oh2018SocialPresenceReview}, potentially influencing their social perceptions of the robot. Second, it may also be explained by the fact that haptic sensation enables realistic physical contact through haptic gloves, which enhances the robot’s perceived social presence, as direct touch fosters a stronger perception of the robot as a socially aware and engaging entity~\cite{Bainbridge2011PhysicallyPresentRobots, Lee2006Tactile}. This type of physical realism has also been shown to intensify emotional engagement in social VR scenarios~\cite{Venkatesan2023CrowdVR}, potentially leading participants to respond to robots in more emotionally authentic ways, which may further influence their affective assessments of the robot. Thus, when assessing human-robot social interactions in VR that involve direct physical contact, the absence of haptic feedback risks undermining the validity of affective-related evaluations.

\subsection{Limitations}
Our study serves as an initial exploration of introducing highly realistic haptic simulation into VR-based HRI studies, which is based on a relatively small sample size with unequal group distributions. Despite careful statistical treatment, this limitation may impact the generalisability of our findings. Second, while our comparison focused on two setups---haptic gloves and hand controllers---the impact of haptic sensation cannot be entirely isolated from other device-related differences, such as equipment weight and varying degrees of hand movement freedom, the latter of which could enhance presence and natural engagement in VR~\cite{krogmeier2023human,Jeremy2003Interpersonal}. Future studies should consider additional control conditions to better disentangle these factors. Finally, while our findings provide valuable insights, the study was conducted within a specific robot help-seeking scenario in urban public spaces. Future research is needed to extend these investigations to a broader range of HRI contexts, incorporating diverse scenarios and interaction instances. 

\section{Conclusion}
Through a comparison of a VR-based human-robot interaction using hand controllers and haptic gloves, we found that enabling haptic sensation resulted in enhanced social and self-presence. Participants also demonstrated more diverse and natural bodily engagement with the robot when haptic sensation, including both tactile and force feedback, was provided. Furthermore, haptic sensation significantly influenced participants' affective-related perceptions of the robot. These findings underscore the importance of incorporating haptic sensations into VR-based HRI studies, particularly for investigating spontaneous reactions to robots and assessing social interactions involving direct physical contact.





\bibliographystyle{IEEETRAN}
\bibliography{sample-base}

\end{document}